\documentclass[12pt,a4paper]{article}
\usepackage{ifpdf}
\ifpdf
\pdfpageattr {/Group << /S /Transparency /I true /CS /DeviceRGB>>}
\fi
\usepackage{jheppub}
\usepackage{amsmath}
\usepackage{caption}
\usepackage{subcaption}

\newcommand{\beq}{\begin{equation}}
\newcommand{\eeq}{\end{equation}}
\newcommand{\bea}{\begin{eqnarray}}
\newcommand{\eea}{\end{eqnarray}}

\newcommand{\nn}{\nonumber}

\begin{document}

\title{A Note on D1-D5 Entropy and Geometric Quantization}
\author[a]{Chethan Krishnan,} \ 
\author[a]{Avinash Raju}
\emailAdd{chethan.krishnan@gmail.com}\emailAdd{avinashraju777@gmail.com}
\affiliation[a]{Center for High Energy Physics, \\
  Indian Institute of Science, \\ 
  Bangalore - 560012, \ \ India}
\keywords{Black Holes in String Theory, D1-D5 system, Fuzzballs}
\abstract{We 
quantize the space of 2-charge fuzzballs in IIB supergravity on $K3$. The resulting entropy precisely matches the D1-D5 black hole entropy, including a specific numerical coefficient. A partial match (ie., a smaller coefficient) was found by Rychkov a decade ago using the Lunin-Mathur subclass of solutions -- we use a simple observation to generalize his approach to the full moduli space of $K3$ fuzzballs, filling a small gap in the literature. 
}

\setcounter{tocdepth}{2}
\maketitle

\section{Introduction and Result}\label{introduction}

Ever since the pioneering work of \cite{Sen, StromingerVafa}, it has been known that string theory can explain the entropy of (at least some classes of) black holes in terms of microstates in an appropriate ensemble. The argument is most under control when the system is BPS, so that one can tune the string coupling to near-zero, while the degeneracy (more precisely an index) remains invariant: the black hole then gets mapped to a system of weakly coupled D-branes whose microstates are easily counted. 

Impressive as it might be, this result is indirect and kinematical. In particular, the  origin of the event horizon in the gravity regime is entirely obscure from the weakly coupled picture. As a corollary, we do not understand various dynamical issues like the information paradox. An outstanding question in this context is this: what happens to the individual microstates as one cranks up the coupling? The fuzzball proposal \cite{Mathur1, Mathur2, BorunAmitabh} is an attempt to answer this question.

The claim of the fuzzball proposal is that at strong coupling, the microstates turn into smooth non-singular solutions of string theory that differ from the black hole solution at the horizon scale, and that the ensemble of these fuzzball microstate ``geometries"\footnote{The word ``geometries" is in quotes because one does not necessarily expect that all of these solutions will be visible in supergravity. However, in the case of the two-charge (D1-D5) black hole, enough fuzzball solutions to capture the leading order entropy are expected in supergravity, this paper (for example) will be an explicit  demonstration of this. However, the question for the 3-charge (D1-D5-p) black hole is much less clear. Large classes of solutions have been constructed (see \cite{superstrata}, and also \cite{hammer}, for the state of the art on this), but it remains to be seen if a finite fraction of the entropy can be found purely in supergravity states. One argument against such a possibility exists in four dimensions due to the existence of pure Higgs states in the weakly coupled D-brane description (see \cite{Dieter}, and the introduction of \cite{SenPioline}). These states do not have a Coulomb branch analogue (and therefore possibly cannot be seen in supergravity) but have a direct interpretation as single-centred black hole microstates -- they are stable under wall-crossing, and have a so-called Lefschetz $SU(2)$ symmetry which can be interpreted as capturing the spherical symmetry of the black hole horizon. In particular, it has been argued in \cite{Sen0903} that {\em all} 4D microstates must have zero angular momentum ($J=0$). In \cite{superstrata}, progress towards the construction of solutions that capture a finite fraction of the entropy of the 5D black hole was made. The solutions have no non-trivial circles, so one does not have contradictions with the arguments of \cite{Sen0903} via dimensional reduction. As an aside -- it is not clear to us if the 4D pure Higgs microstates are forbidden (or not) from having a supergravity description in terms of some hitherto undiscovered $J=0$ microstates. 
In any event, our goal here is not to get into the debate on what fraction of the 3-charge microstates can be seen in supergravity. Our goals are modest and limited to the 2-charge system. But we feel it is necessary to give the reader some context regarding the status of SUGRA fuzzball microstates in the interest of transparency.} is what the black hole is comprised of. For the two charge (D1-D5) black hole in Type IIB string theory on $K3$ or $T^4$ there are various arguments that the entropy of the system at leading order can be reproduced entirely via fuzzball solutions that are visible within the supergravity description as smooth horizonless solutions \cite{SkenderisReview}. One argument is to quantize the phase space of fuzzball solutions and counting the number of states to see whether it reproduces the D1-D5 entropy. This was done for the original Lunin-Mathur subclass of 2-charge fuzzball solutions \cite{LuninMathur} by Rychkov \cite{Rychkov}. But Rychkov's result, 
\bea
S \approx 2 \pi \sqrt{\frac{2}{3} N_1 N_5} \label{RychkovEntropy}
\eea
does not reproduce the full entropy, which is 
\bea
S \approx 4 \pi \sqrt{ N_1 N_5} \label{K3Entropy}
\eea
for the $K3$ black hole\footnote{We will not consider the $T^4$ compactifications in this paper for reasons explained in the final section.}. This is not surprising because the approach of Rychkov \cite{Rychkov} did not incorporate the complete family of 2-charge fuzzballs, in particular they do not include the excitations in the compact directions. The result is nonetheless suggestive because it does capture the correct scaling of the entropy with the D1 and D5 charges. 

In this paper, we will consider the complete phase space of 2-charge fuzzball solutions for the specific case of compactification on $K3$. These were constructed by Kanitscheider, Skenderis and Taylor (KST) \cite{Skenderis}. We will use geometric quantization of the phase space of those solutions. We will be able to argue that a simple generalization of the Rychkov symplectic form is the correct choice on them, enabling us to extend his result to the full phase space. We find that the result (\ref{K3Entropy}) 
is precisely reproduced, giving closure to a gap in the literature.


\section{General Two Charge Fuzzballs on $K3$}

The general two-charge fuzzballs for the $K3$ case were constructed in \cite{Skenderis}, we will refer to them as the KST fuzzball solutions. We follow the conventions in \cite{SkenderisReview}:
\bea \label{genK3fuzzball}
   ds_{\rm string}^2 &=& \frac{f_1^{1/2}}{\tilde{f_1} f_5^{1/2} }[-(dt-A_i
     dx^i)^2+(dz-B_i dx^i)^2] + f_1^{1/2} f_5^{1/2} dx_i dx^i
+  f_1^{1/2} f_{5}^{-1/2} ds^2(K3),  \nonumber \\
   e^{2\Phi} &=& \frac{f_1^2}{f_5 \tilde{f_1} }, \qquad B^{(2)}_{tz} =
   \frac{{\cal A}}{f_5 \tilde{f}_1},
\qquad B^{(2)}_{\bar{\mu} i} = \frac{{\cal A} {\cal B}^{\bar{\mu}}_i}{f_5 \tilde{f_1} }, \\
C^{(0)}&=& - f_{1}^{-1}{\cal A}, \qquad B^{(2)}_{ij} = \l_{ij} +
\frac{2 {\cal A} A_{[i} B_{j]}} {f_5 \tilde{f}_1 }, \qquad
B^{(2)}_{\rho\sigma} = f_{5}^{-1} k^{\gamma}
\omega_{\rho \sigma}^{\gamma},  \nn \\
   C^{(4)}_{tzij} &=& \l_{ij} + \frac{{\cal A}}{f_5 \tilde{f_1}}(c_{ij} + 2 A_{[i}B_{j]}),
\qquad C^{(4)}_{\bar{\mu}ijk} = \frac{3 {\cal A}}{f_5 \tilde{f}_1}
{\cal B}^{\bar{\mu}}_{[i}c_{jk]}, \nn \\
   C^{(4)}_{tz \rho \sigma}
&=& f_{5}^{-1} k^{\gamma} \omega^{\gamma}_{\rho \sigma},
\qquad C^{(4)}_{ij \rho \sigma} = (\l^{\gamma}_{ij} +
f_{5}^{-1} k^{\gamma}c_{ij})\omega^{\gamma}_{\rho \sigma}, \qquad
   C^{(4)}_{\rho \sigma \tau \pi} = f_{5}^{-1} \cal A \epsilon_{\rho
     \sigma \tau \pi}, \nn \\
     C^{(2)}_{tz} &=& 1-\tilde{f}_1^{-1}, \qquad C^{(2)}_{\bar{\mu}i} = -
   \tilde{f}_1^{-1} {\cal B}^{\bar{\mu}}_i, \qquad C^{(2)}_{ij} = c_{ij} - 2 \tilde{f}_1^{-1}
A_{[i}B_{j]}. \nn 
\eea
The metric is in the string frame. The  
$\omega^{\gamma} \equiv (\omega^{\alpha_+}, \omega^{\alpha_-})$ are a basis of self-dual and anti-self-dual 2-forms on $K3$ with $\gamma =
1,\cdots,22$ where 22 is the second Betti number of $K3$. The labels\footnote{We stick to the $\alpha_{\pm}$ notation that is used in \cite{SkenderisReview} because it can be adapted to the $K3$ case as well.} take the values  $\alpha_{+} = 1,2,3$ and  $\alpha_{-} = 1,\cdots 19$. The intersection numbers of the forms are
\bea\label{K3intersec}
   d_{\gamma\delta} = \frac{1}{(2 \pi)^4 V} \int_{K3} \omega_2^{\gamma} \wedge \omega_2^{\delta}.
\eea
The integration constant in $C_{tz}^{(2)}$ ensures that the potential vanishes at infinity -- the solutions depend on the harmonic functions
$(H,K,A_{i},{\cal A}, {\cal A}^{\alpha_-})$ via
\bea
\label{D1D5K3aux}
f_{5} &=& H, \qquad \tilde{f}_1 = 1 + K - H^{-1} ({\cal A}^2 + {\cal A}^{\alpha_-} {\cal A}^{\alpha_-}), \qquad
f_1 = \tilde{f}_1 + H^{-1} {\cal A}^2, \nn \\
d\l^{\gamma} &=& *_4 dk^{\gamma}, \qquad d\l = *_4 d {\cal A}, \qquad
{\cal B}^{\bar{\mu}}_i = (-B_i,A_i), \\
k^{\gamma} &=& (0_3, \sqrt{2} {\cal A}^{\alpha_-}),
\qquad dB = -*_4 dA, \qquad
   dc = - *_4 df_5. \nn
\eea
Here $\bar{\mu} = (t,z)$ and $*_{4}$ is the Hodge dual over flat
$R^4$. The Hodge dual in the Calabi-Yau $K3$ metric is given by $\epsilon_{\rho \sigma \tau \pi}$.

The solutions in \cite{Skenderis} correspond to the choice of Harmonic functions given by 
\bea
H &=& 1 + \frac{Q_5}{L} \int_0^L \frac{dv}{|x-F(v)|^2},
\qquad A_i = -\frac{Q_5}{L}\int_0^L \frac{dv \dot{F}_i(v)}{|x-F(v)|^2}, \nn \\
&& \hspace{0.2in} K = \frac{Q_5}{L} \int_0^L \frac{dv (\dot{F}(v)^2 +
\dot{\cal F}(v)^2 + \dot{\cal{F}}^{\alpha -}(v)^2 )}{|x-F(v)|^2}, 
\label{harmonic} \\
   {\cal A} &=& -\frac{Q_5}{L}\int_0^L \frac{dv
     \dot{\cal F} (v)}{|x-F(v)|^2}, \qquad
{\cal A}^{\alpha_-} = -\frac{Q_5}{L} \int_0^L \frac{dv \dot{\cal{F}}^{\alpha_-}(v)}{|x-F(v)|^2}. \nn 
\eea
Here, $|x-F(v)|^2$ is to be understood as $\sum_i|x^i-F^i(v)|^2$, we will often suppress summation over the index $i$. The 5-brane charge $Q_5$ and the length of the defining curve $L$ in the D1-D5 system (see \cite{SkenderisReview}) are related through the radius of the $z$-circle $R$  via
by
\bea
L = 2 \pi Q_5/R.
\eea
A relation that is useful and important for us is the expression for the D1 charge $Q_1$:
\bea \label{charges}
Q_1 = \frac{Q_5}{L} \int_0^L dv (\dot{F}^i(v)^2 +
\dot{\cal F}(v)^2 + \dot{\cal{F}}^{\alpha_-}(v)^2 ).
\eea
The integral charges are given by
\bea \label{q1}
Q_5 = g_s \alpha' N_5, \ \ Q_1 = g_s \frac{N_1 (\alpha')^3}{V}.
\eea
Here $(2 \pi)^4 V$ is the volume of $K3$. Henceforth, we will set $\alpha'$ to unity.

The Lunin-Mathur solutions correspond to setting ${\cal F}(v)=0={\cal{F}}^{\alpha_-}(v)$ in the KST fuzzballs. The detailed form of the KST solution will not be necessary to follow most of our discussions, but we present it here for two reasons: 
\begin{itemize}
\item In checking (\ref{KSTHamiltonian}), which is the key observation of this paper, from first principles, we {\em will} need the details of the solution. 
\item We want to emphasize that at least superficially, the general fuzzballs are substantially more complicated than the Lunin-Mathur fuzzballs \cite{LuninMathur}.
\end{itemize}

\section{The Consistent Symplectic Form}

The basic idea of geometric quantization is to quantize the phase space, count the states in the Hilbert space and use that as the definition of the micro-canonical entropy.  The phase space and the space of solutions have a one-to-one map, so we can also work with the latter. The goal then is to compute the symplectic form on the space of solutions and then quantization can proceed as usual. In principle this is straightforward, but it is bound to be a complicated problem for the fuzzball solutions presented in the last section. 

Indeed, even for the Lunin-Mathur solutions the task was complicated, and Rychkov used two simplifying facts to make the problem tractable, and to compute the restriction of the full IIB supergravity symplectic form\footnote{We will not write down the full IIB symplectic form, it can be found in many of the references we have already listed.} onto the moduli space of solutions. The first was that the the Lunin-Mathur solutions are time-independent, which is a fact that is trivially true for our more general KST solutions as well. The second was that the Hamiltonian, when restricted to the moduli space took a specific simple form \cite{Rychkov}:
\bea
H|_{{\cal M}_{LM}}=\frac{R V}{g_s^2}\left(\frac{Q_5}{L}\int_0^L \dot{F}^i(v)^2 \ dv + Q_5\right), \label{LMHamiltonian}
\eea 
where the subscript $LM$ on the left hand side denotes the fact that we are working with the Lunin-Mathur subclass of solutions. Using these facts, it was argued in pages 7-8 of \cite{Rychkov} that the symplectic form should take the form
\bea
\Omega=\frac{1}{2 \alpha} \int \delta\dot{F}^i(v) \wedge \delta F^i (v) dv \label{RychkovOmega}
\eea
where $\alpha$ can only depend on the various integrals of motion determined by the curve functions $F^i(v)$:
\bea
\alpha \equiv \alpha\left[ \int \dot F^i(v)^2 dv,   \int \ddot{F}^i(v)^2 dv, . . . \right] \label{alphagen}
\eea
Furthermore, (a) by computing the symplectic form explicitly from the IIB symplectic form for a subclass of curves with chosen $F^i(v)$, and (b) finding in that class of curves that $\alpha = \pi \mu^2$ is a numerical constant\footnote{Here, $\mu^2=\frac{g_s^2}{R^2 V}$.}, Rychkov argued \cite{Rychkov} that the only expression of the form (\ref{alphagen}) which can reduce to such a constant on the subclass of curves, is the choice $\alpha=\pi \mu^2$ on the {\em entire} Lunin-Mathur moduli space. This fixed the symplectic form for the Lunin-Mathur fuzzballs, allowing a direct determination of the entropy of those solutions by geometric quantization.

At first sight, the generalization from Lunin-Mathur to KST fuzzballs seems formidable. The solution is substantially more complex, and for the $K3$ case, there are 20 (=19+1) new independent functions in the solution now. It is also clear, that the Hamiltonian of the KST fuzzballs must be different from (\ref{LMHamiltonian})\footnote{Indeed this is necessary, if one has hopes of reproducing the full entropy by doing geometric quantization.}. Despite these potential complications, we will show in this paper that the IIB supergravity Hamiltonian, when restricted to the KST solutions, retains enough of the simple features of the Lunin-Mathur solutions that we can adapt the Rychkov arguments to get the complete answer without getting bogged down in the details. 

The basic observation is that the energy in the KST case can be directly computed, and it takes the simple form 
\bea
H|_{{\cal M}_{KST}}=\frac{R V}{g_s^2}\left(\frac{Q_5}{L}\int_0^L  (\dot{F}^i(v)^2 +
\dot{\cal F}(v)^2 + \dot{\cal{F}}^{\alpha_-}(v)^2 )\ dv + Q_5\right). \label{KSTHamiltonian}
\eea
despite the added complexity of the KST solutions\footnote{The subscript $KST$ on the left hand side denotes  that we are working with the full family of KST fuzzballs.}.  This can be obtained straightforwardly via the ADM approach (we sketch it in an appendix), but it is easy to convince oneself that this answer is as it should be, as follows -- Using (\ref{charges}) and (\ref{KSTHamiltonian}) we can show immediately that the total mass of the system is
\bea
E_{\rm brane-mass}=\frac{N_1 R}{g_s} + \frac{N_5 R V}{g_s},
\eea
and since the system is BPS, this is something we would expect\footnote{In fact, it was this observation that lead us to first guess that the answer for the Hamiltonian might be simple. Once having reproduced the correct entropy using the guess, one can also do the direct computation of (\ref{KSTHamiltonian}), see Appendix.}. Now, the form (\ref{KSTHamiltonian}) is very closely related to the original energy functional in Rychkov's computation (\ref{LMHamiltonian}), with the crucial fact that all the independent functions enter democratically and quadratically in it. In effect, therefore the arguments leading to the symplectic form (\ref{RychkovOmega}) in \cite{Rychkov} go through exactly as before, with the only new ingredient that it should also involve terms from the new functions:
\bea
\Omega=\frac{1}{2 \alpha} \int (\delta\dot{F}^i(v) \wedge \delta F^i (v)+\delta\dot{{\cal F}}(v) \wedge \delta {\cal F} (v)+\delta\dot{\cal{F}}^{\alpha_-}(v) \wedge \delta {\cal F}^{\alpha_-} (v)) dv. \label{OurOmega}
\eea 
Now, since the subclass of curves considered in \cite{Rychkov} to argue that $\alpha$ must be the numerical constant $\pi \mu^2$ is {\em also} a subclass of the curves considered here, it immediately follows that the $\alpha = \pi \mu^2$ here as well. This fixes the symplectic form for the KST solutions completely. 

\section{Entropy Match from Quantized Phase Space}

Once we have the symplectic form, we have everything we need to quantize and compute the entropy. 
Since all the curve functions enter democratically in the discussion, we will define
\bea
F^I(v) \equiv (F^i (v), {\cal F} (v), {\cal F}^{\alpha_-} (v)).
\eea
Note that $I$ takes 24 values because of this definition. Now, the standard approach is to expand the curve functions into Fourier oscillators and to count the modes, see section 2 of \cite{Rychkov} for a clear discussion. The only difference between there and here is that there the indices $i$ in $F^i(v)$ took only four values reproducing a result that would be equal to that of four chiral bosons (ie., central charge $c=4$). Here we get the analogue of 24 chiral bosons ($c=24$). The latter is what is indeed expected for the D1D5 black hole on $K3$, see eg. p.28 of \cite{SkenderisReview}. The answer can therefore be obtained via the Cardy formula
\bea
S \sim 2\pi \sqrt{\frac{c}{6}N_1 N_5} = 4 \pi \sqrt{N_1 N_5},
\eea
reproducing the classical Bekenstein-Hawking entropy of the $K3$ hole.

\section{Comments}

We have found that the geometric quantization of the general fuzzball moduli space on $K3$ reproduces the corresponding D1-D5 entropy on the nose. The possibility that a more complicated structure for the symplectic form could arise and complicate the computation has been raised in the literature (see discussion after eqn (4.76) in \cite{SkenderisReview}), but by working with the energy of the general fuzzballs we have shown that the problem can be solved by a simple generalization of the Rychkov argument. The final symplectic form is indeed simple and democratic in all the curve functions. With the malice of hindsight, perhaps one could have taken the existence of 24 unknown functions in the KST solutions as a hint of this, already at the time they were constructed \cite{Skenderis, SkenderisReview}. 

It will be interesting to repeat a similar computation in the $T^4$ case. However, unlike in the $K3$ case, the solution (\ref{genK3fuzzball}) in the $T^4$ case (ie., now $\alpha_-$ only in the range $\alpha_-=1,2,3$ corresponding to the anti-self dual 2-forms on the 4-torus) does not describe the most general fuzzball solution \cite{SkenderisReview}. It describes only the bosonic excitations, one needs to further add fermionic excitations. Related questions seem to have been addressed in \cite{Marika}, 
we hope to come back to the $T^4$ computation sometime in the future. 

\acknowledgments
We thank Iosif Bena, Stefano Giusto, Shiraz Minwalla, Rodolfo Russo, Joan Simon, Dieter Van den Bleeken, Amitabh Virmani and Nick Warner for discussions on general fuzzball related matters, and Borun Chowdhury and Kostas Skenderis for discussions on the specifics of this paper. We re-thank Kostas for comments on a previous version of the manuscript. CK thanks the School of Mathematics, University of Edinburgh and Swansea University, for hospitality.\\

\appendix\section*{Appendix: ADM Mass from 5D Reduction}

In this appendix we will briefly sketch how to get (\ref{KSTHamiltonian}) directly, without using the BPS argument. In \cite{Rychkov} the analogous result is obtained via the formula for asymptotic charges in general relativity. We will get our result by reading off the fall-offs of the $g^{Einstein5D}_{tt}$ piece of the effective five dimensional metric in the Einstein frame and identifying its ADM energy. This reproduces the result of \cite{Rychkov} when restricted to the Lunin-Mathur subclass of solutions. 

We want to view the KST solutions as five dimensional asymptotically flat solutions. So we wish to obtain the effectively five dimensional metric that captures the string frame KST metric presented in Section 2. The KST solution falls into the standard Kaluza-Klein ansatz when thought of as a 10 D solution. See for example Appendix E of Kiritsis \cite{Kiritsis}, whose notations we follow. The reduction there is done starting with the string-frame metric, which is exactly what we want. From eqns (E.3-E.4) in \cite{Kiritsis} and the structure of KST solution (specifically, the dilaton and the metric components in the compact directions), one can see that the reduction of the 10D string-frame KST metric gives rise to a 5D string(-like)-frame metric and a 5D ``dilaton" (this is the field $\phi$ defined in eqn. E.4 of \cite{Kiritsis}). The latter can be computed to be
\bea
\phi= \frac{3}{8} \ln f_1 -\frac{1}{4} \ln \tilde f_1 + \frac{1}{8} \ln f_5
\eea
The effective five dimensional Einstein frame metric can then be obtained via a conformal rescaling
\bea
g^{Einstein5D}_{tt}=e^{-4 \phi /3} \ g_{tt}
\eea
where $g_{tt} \sim \frac{f_1^{1/2}}{\tilde f_1 f_5^{1/2}}$ is the string-frame metric component. Explicitly, this yields
\bea
g^{Einstein5D}_{tt} \sim \frac{1}{(\tilde f_1 f_5)^{2/3}}.
\eea
Now it is straightforward to read off the ADM energy from the subleading fall-off of this metric component, and the result (\ref{KSTHamiltonian}) follows. Eqn. (2.16) of \cite{Peet} is useful for fixing ADM conventions when comparing fall-offs. Note that $|x-F(v)|^2$ can be approximated $|x|^2 \sim r^2$ (where $r$ is the radial coordinate in 4+1 D) and taken outside the integral to the order that is relevant for calculating the ADM mass.


\bibliographystyle{JHEP}
\bibliography{d1d5.bib}

\end{document}